# The Ripple Effect of Retraction on an Author's Collaboration Network

Kiran Sharma
School of Engineering and Technology,
BML Munjal University, Gurugram, India
E-mail: kiran.sharma@bmu.edu.in

Satyam Mukherjee *
School of Management and Entrepreneurship,
Shiv Nadar Institution of Eminence, Greater Noida, India
E-mail: satyam.mukerjee@gmail.com

* Corresponding author

**Abstract**

Scientists involved in scientific misconduct may face social stigmatization, leading to isolation and limited opportunities for collaboration. The reputation of every individual is reflected on the team, as the fraud attempted by any member will be reflected on the team. Earlier studies pointed out the impact of citation penalty on the prior work of coauthors, the effect of retraction on a co-author's research career, and stigmatization through mere association. This paper explores the formation and dynamics of the networks of authors who faced retractions and their "innocent coauthors" who never faced retractions in their careers. Leveraging a dataset of 5972 retracted papers involving 24209 authors, we investigate whether scientific misconduct reduces collaborative ties of misconducting authors as opposed to those who never faced allegations of academic misconduct. We observe that the network structure of authors involved in retractions does not change significantly over the years compared to that of the "innocent coauthors". Our results suggest that stigmatization rarely affects the collaboration network of stigmatized authors. Our findings have implications for institutions adopting stringent measures and fostering ethical practices research.

## 1 Introduction

Fraud and trust are important scientific considerations (Fuchs & Westervelt, 1996). Most scientists conduct their research with integrity and adhere to ethical standards. However, instances of scientific fraud do occur, and they can have significant consequences for the scientific community and public trust. Research misconduct refers to the violation of ethical standards and principles in the conduct of scientific research. It refers to intentional deception in the research process, such as fabricating or falsifying data, plagiarizing others' work, or misrepresenting research findings. Instances of fraud can undermine the credibility of scientific research and erode public expectations of science.

Even though statistics suggest that retractions of papers are relatively rare, the recent surge in retracted publications has led scientific gatekeepers to filter out "bad science" (Brainard & You, 2018). Various factors contribute to researchers engaging in misconduct and understanding them fully is still limited. Some scholars propose that the pressure to "publish or perish" in a competitive research environment is the key reason (Casadevall et al., 2014; Qiu, 2010). In the publish-or-perish culture of academia, where researchers are evaluated based on their publication records and ability to secure research funding, there can be intense pressure to produce positive results and publish frequently. The pressure to publish can create an environment that promotes unethical practices (Kiai, 2019).

Replicating and verifying research findings by other scientists is crucial to validate the results and detect any potential fraud (McNutt, 2014). Openness in scientific research, such as sharing data, methodologies, and results, promotes transparency and allows for scrutiny and replication by others. Open data initiatives can help prevent fraud by facilitating independent analysis and verification. Replication studies enhance the robustness and reliability of scientific findings. However, insufficient reporting and lack of transparency in scientific publications can impede replication efforts (Baker, 2015, 2016). Moreover, replication studies often require significant resources, including funding, time, and access to specialized equipment or facilities (Poldrack, 2019).

Engaging in scientific misconduct can severely damage a researcher's credibility within the scientific community (Azoulay et al., 2017). Researchers involved in scientific misconduct may experience significant damage to their professional reputations (Mongeon & Larivière, 2016). When misconduct is uncovered, it can create a sense of betrayal and erode the trust that is crucial for successful collaboration. This relationship damage can hinder future collaborations and networking opportunities (Hussinger & Pellens, 2019; Jin et al., 2019). Earlier studies on the effect of retraction on scientific careers posit the citation penalty on the prior work of coauthors (Jin et al., 2019), the effect of retraction on a co-author's research career (Mongeon & Larivière, 2016) and stigmatization through mere association (Hussinger & Pellens, 2019).

Scientists involved in scientific misconduct may face social stigmatization, leading to isolation and limited opportunities for collaboration (Goldstein & Johnson, 1997). Stigma by association in science refers to the negative perceptions or judgments that can be placed on individuals or groups based on their association with someone involved in scientific misconduct. Losing trust and credibility can incur financial costs to funding agencies and jeopardize academic job opportunities (Stern et al., 2014).

While prior research on the effect of scientific fraud on a co-author's career has dealt with reduced performance of "innocent co-authors" post-retraction in terms of publication (Mongeon & Larivière, 2016; Shuai et al., 2017) or decline in citations (Jin et al., 2019), they have ignored the network structure of collaborative links for "guilty" and "innocent" authors. This paper addresses this gap by exploring the formation and dynamics of the networks of authors who faced retractions and their "innocent coauthors" who never faced retractions in their careers.

Specifically, we investigate whether scientific misconduct reduces collaborative ties of misconducting authors as opposed to those who never faced allegations of academic misconduct. We describe the methodology, case setting and descriptive insights of the data in Section 2. Section 3 covers the exploratory findings of our empirical research, followed by a discussion in Section 4. We conclude in Section 5.

## 2 Methodology

To obtain our sample of retracted papers, we explored the entire research career of an individual. First, from the Web of Science database, we extracted 5972 papers through 2020 with the document type "Retracted Publication". Second, we mapped all 5972 papers with the Scopus database to locate unique authors from all retracted publications. Scopus filtered 24209 authors with unique identifiers out of 5972 retracted papers. These authors are marked as "Retracted authors", i.e., authors who received at least one retraction in their career.

Next, for each of those 24209 authors, their entire research career and collaboration were explored in terms of publication year, co-authors ID, and paper DOI. For all 24209 authors, we got a total of 822762 publications till 2020. Finally, we extracted the information of the co-authors of retracted authors, which yielded 2144425 such authors who collaborated with retracted authors at some point in their careers and never faced retraction.

### 2.1 Data Description

This section provides a descriptive analysis of the data on retracted publications.

### 2.1.1 *Cumulative distributions*

Following the work on one of the first publications on academic misconduct and retracted papers (Fang et al., 2012), we estimate the time-to-retraction as the difference between retraction and publication years. Figure 1 shows the empirical cumulative distribution of the time-to-retraction in the semi-log y-axis. The distribution is broad and has a slow decay, with the maximum time-to-retraction value being 30 years. The slow decay shows that 90% of the papers are retracted within the first 10 years of their publications. It is also evident that most papers are not retracted immediately after publication, suggesting they were retracted after scrutiny and failing reproducibility by peer groups.

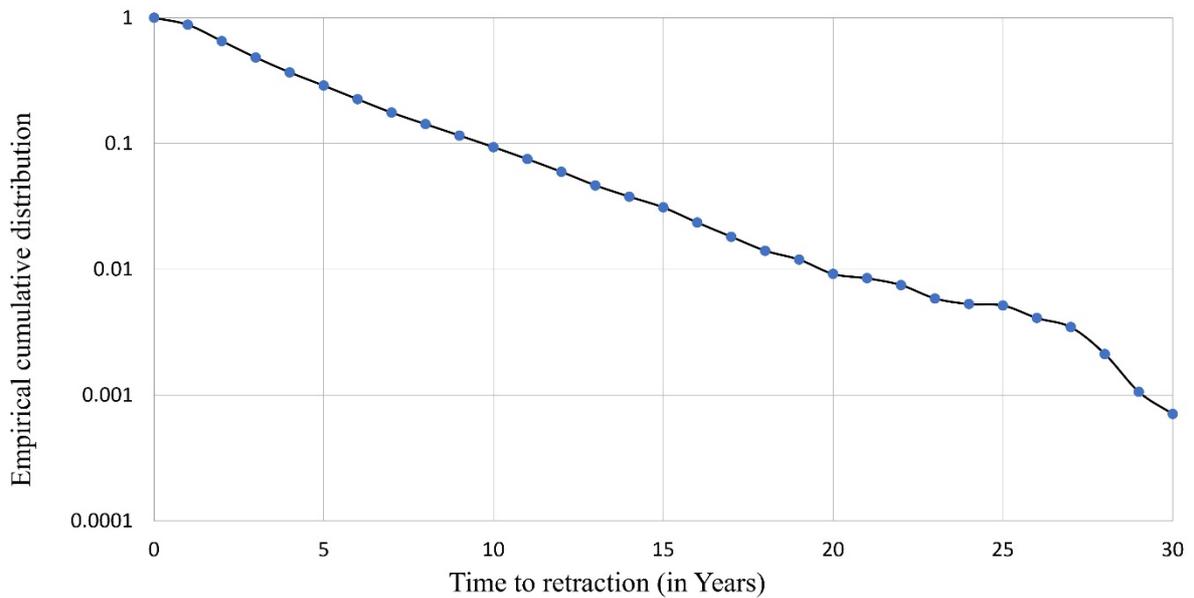

**Figure 1.** Distribution of difference in retraction year and publication year of a retracted paper

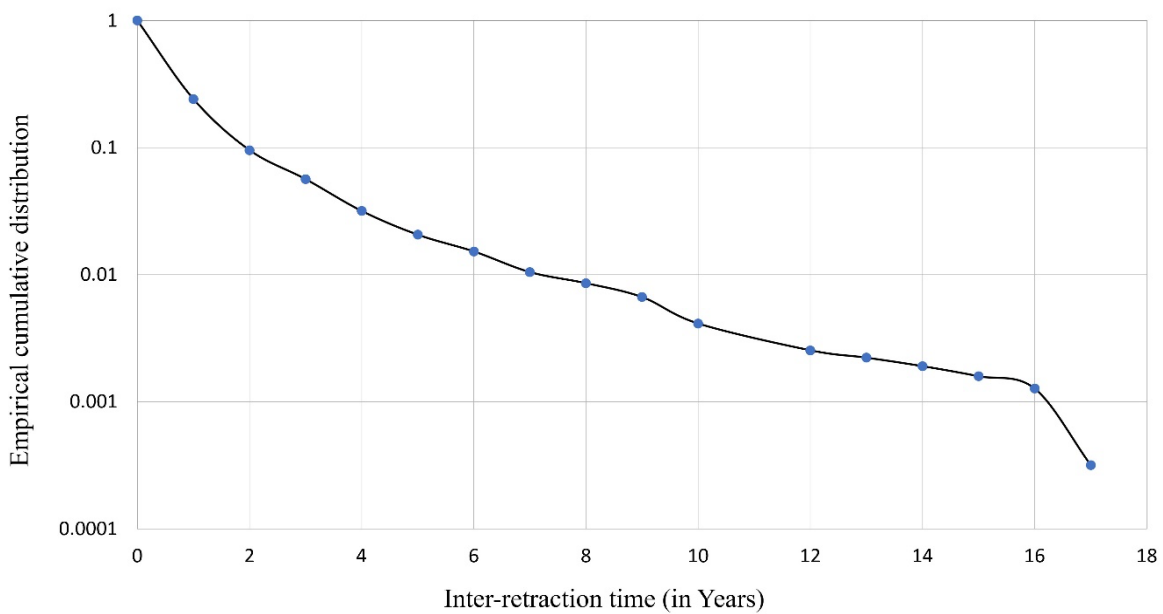

**Figure 2.** Distribution of inter-retraction times

Next, we investigate the inter-retraction patterns of the retracted publications. It is evident that there are authors responsible for multiple retractions. For each author, we chose the set of retracted papers and their year of retraction. Next, for every author, we estimate the inter-retraction times (in years) as the difference in years of successive retractions. Figure 2 shows the empirical cumulative distribution of all inter-retraction times aggregated over all the authors in the data — the distribution decays rather faster, with the maximum value of inter-retraction being 17 years. The distribution shows that 75% of the authors have multiple retractions within the first year of retraction. Overall, the average inter-retraction time is 0.5 years, suggesting guilty authors face multiple retractions quickly.

*2.1.2 Network of Collaboration*

In Figure 3, we provide a visual summary of the collaboration network of the authors for papers published in 1976. As mentioned in the Methodology, we categorized the authors with at least one retraction in their career as "Got Retraction", while authors who never faced retractions are categorized as "Never Retracted". The "Never Retracted" authors are those authors who collaborated with the guilty authors at some point in their careers, but those publications never faced allegations of any scientific misconduct. For each publication year, we created the collaboration network of authors, the weight of the links being estimated as the frequency of collaborations for that publication year. This yields a weighted and undirected collaborative network of authors classified into two categories, as mentioned above.

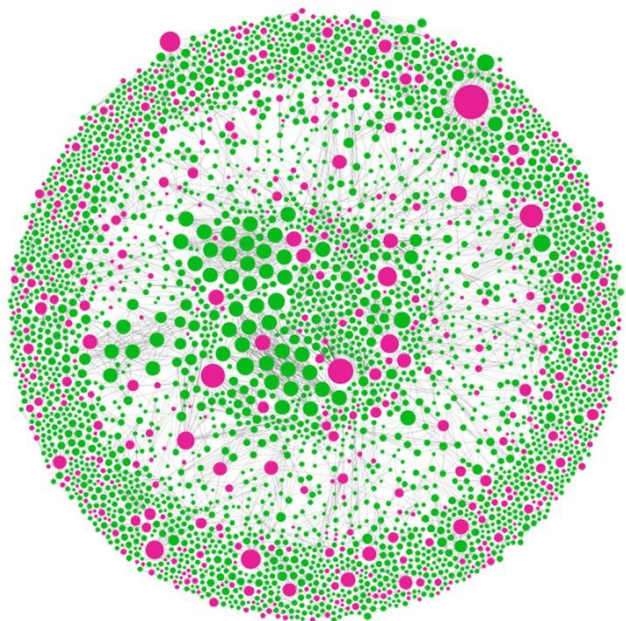

**Figure 3.** Network of collaboration of authors with at least one retraction in their career (dark-pink) and authors with no retraction (green). The node's size is the strength (weighted degree), and the weight of the links is the number of times two authors collaborated. The network consists of 3363 nodes and 7676 edges. In this figure, we consider papers published in 1976.

One of the fundamental metrics used in network science is the degree of a node, defined as the number of direct neighbors. In the context of collaboration networks, the degree of an author measures the number of direct collaborators. Extending the concept of degree to its weighted version, we define the strength ($s_i$) of an author $i$ as:

$$s_i = \sum_{j=1}^{N} A_{ij} w_{ij}$$

Where $w_{ij}$ is the weight of a collaborative link estimated as the frequency of collaborations between authors $i$ and $j$. The adjacency matrix $A_{ij}$ takes a value of 1 if there is a collaborative link between $i$ and $j$; 0 otherwise. Typically, authors with higher values of degree and strength would have higher frequencies of collaborative ties.

We also quantify the collaboration networks' local density by evaluating the authors' local clustering coefficient. The local clustering coefficient (LCC) is estimated as the fraction of the number of collaborative links to the number of possible collaborative links between the neighboring authors of an author. Accounting for the weight of the collaborative links, the weighted local clustering coefficient ($C_i^w$) of an author $i$ is defined as:

$$C_i^w = \frac{1}{s_i(k_i - 1)} \sum_{jh} \frac{(w_{ij} + w_{jh})}{2} A_{ij} A_{jh} A_{hi}$$

Where the symbols have their usual meanings, with $k_i$ being the degree of an author $i$. Higher values of weighted LCC imply that interconnected triplets are more likely to form between authors with greater collaborations. Next, we compare these two network metrics for "Got Retraction" and "Never Retracted" authors.

## 3  Results

First, we statistically determine whether there is any significant difference between the two groups of authors in terms of the network metrics. Implementing the Wilcoxon rank-sum test, we observe that a significant difference exists between the two groups of authors – degree ($z$ = 87.074; *p*-value *<0.001*), strength ($z$ = 258.386; *p*-value *<0.001*), LCC ($z$ = -697.153; *p*-value *<0.001*), and weighted LCC ($z$ = -42.269; *p*-value *<0.001*). We also observe that the average degree of authors who "Got Retraction" is 24.89 (std dev = 41.19), significantly higher than "Never Retracted" authors with an average degree of 19.81 (std dev = 32.59). A similar pattern is observed for the strength of authors in the two groups (Got Retraction: average = 79.96, std dev = 278.61; Never Retracted: average = 45.26, std dev = 290.02).

We observe a contrasting pattern while comparing the LCC and weighted LCC for authors in "Got Retraction" and those in "Never Retracted". The average LCC for "Never Retracted" authors is 0.91 (std dev = 0.19), significantly higher than that of "Got Retraction" authors (average LCC = 0.54; std dev = 0.33). Interestingly, the pattern is changed when we compare the weighted LCC for the two groups (Got Retraction: average = 0.013, std dev = 0.023; Never Retracted: average = 0.010, std dev = 0.014).

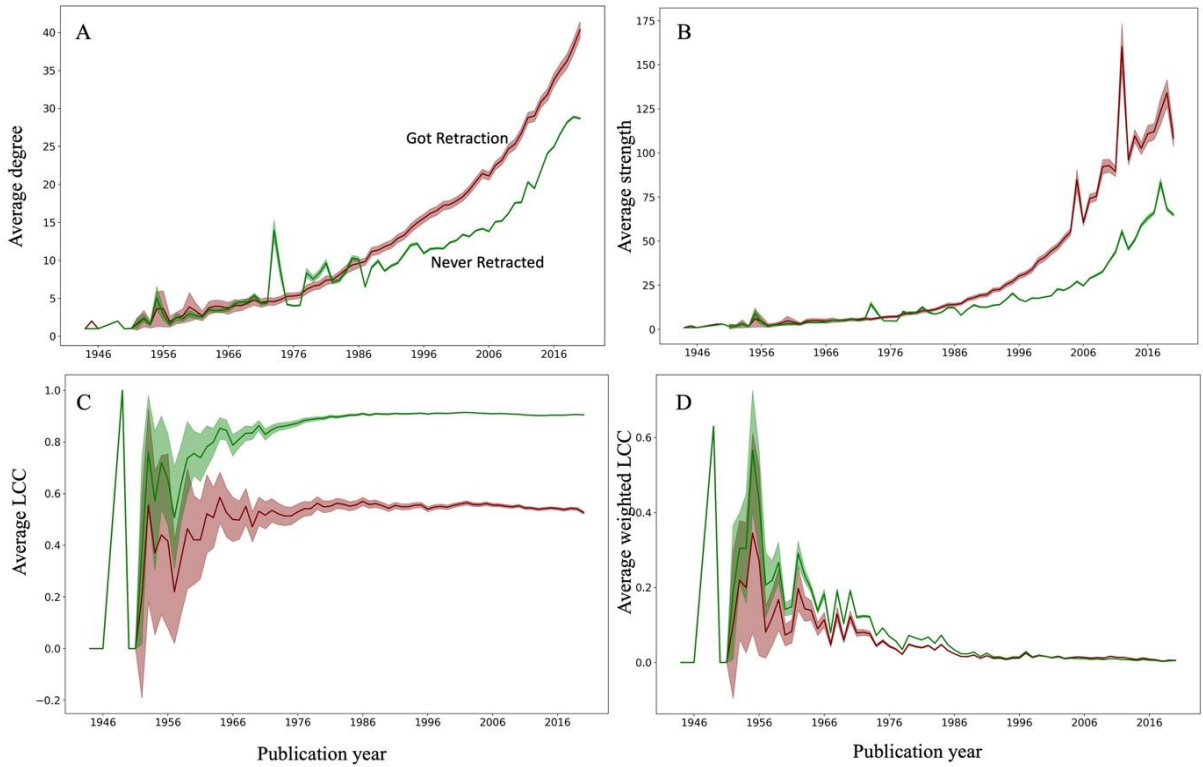

**Figure 4.** Average (A) degree, (B) strength, (C) local clustering coefficient (LCC), and (D) weighted LCC of "Got Retraction" (Maroon color) and "Never Retracted" (Green color) authors with publication year.

We explore the temporal trends of the network metrics with publication year for the two categories of authors. Focusing on the degree of the authors from the two groups, we notice no significant difference between the two groups in the initial transient period of publication years through 1970 (See Figure 4A). Even though the increase in the average degree of authors is slow between 1944 and 1970, the upward trend indicates there is not much stigmatic effect on misconducting authors. With the surge in publications over the years, the authors' average degree increases faster. We observe that the "Got Retraction" authors have consistently higher average degrees than the "Never Retracted" authors. Focusing on the strength of the authors (See Figure 4B), we again observe the same pattern as seen in Figure 4A.

The unweighted and weighted local clustering coefficients (LCC) for the two groups of authors show contrasting results (See Figure 4C-4D). As observed above, during the initial transient periods of the 1940s, the average LCC kept increasing through 1970. We observe that the average LCC for "Never Retracted" authors is significantly higher than the other misconducting group of authors ("Got Retraction"). Interestingly, post-1975, the values of LCC for the two groups of authors stabilize around mean values of 0.91 ("Never Retracted") and 0.54 ("Got Retraction"). The significantly higher values of LCC for "Never Retracted" authors suggest that collaboration among the co-authors of misconducting authors is significantly diminished. A contrasting picture is observed for the weighted LCC (See Figure 4D), where the values keep decreasing over time and flatten around a mean value of ~ 0.01. Further, the values of weighted LCC for the two groups appear to be relatively close since 2010. Figures 4C-4D suggest that co-authors of both groups of authors face similar consequences.

## 4 Discussions

Retractions can result in significant repercussions for authors' professional trajectories such as damage to reputation (Hussinger & Pellens, 2019), loss of trust (Fuchs & Westervelt, 1996), impact on career progression (Azoulay et al., 2017), stigmatization (Goldstein & Johnson, 1997), limited publishing and funding opportunities, and collateral impact (Mongeon & Larivière, 2016) which may lead to their exit from the realm of the scientific community (Jin et al., 2019). It's crucial to acknowledge that while certain retractions stem from unethical conduct or deliberate misdeeds, others may arise from genuine errors, methodological shortcomings, or unexpected mistakes. There exists a considerable body of literature dedicated to categorizing such distinctions (Casadevall et al., 2014; Fang et al., 2012). This paper investigated the collaborative ties among authors who have faced and not faced allegations of scientific misconduct (retracted paper) indexed in the Web of Science and Scopus.

Contrary to what we initially anticipated, our observations reveal that authors engaging in misconduct and facing retractions do not experience significant consequences regarding their collaborative relationships. Prior studies have shown that retractions don't affect collaboration practices among co-authors, at least in the biomedical field (Mongeon & Larivière, 2016). However, our findings on the unweighted and weighted local clustering coefficients portray a different story. Even though new collaborations among the co-authors of "Got Retraction" are not forming, the frequency of existing collaboration persists.

Our results suggest that stigmatization rarely affects the stigmatized authors (Hussinger & Pellens, 2019). This fortuitous finding could be attributed to several factors. First, the nature of collaborative ties might not be heavily influenced by a single instance of misconduct or retraction. Collaborative partnerships are often built on a combination of factors, including shared research interests, expertise, and prior successful collaborations. An isolated retraction incident might not be sufficient to sever these ties, especially if the misconduct is not directly related to the collaborative work.

Second, the scientific community's response to misconduct and retractions can vary. Some cases might attract considerable attention and scrutiny, leading to damage to the researchers' reputation and collaboration opportunities. However, the misconduct might not receive widespread attention in other instances, allowing authors to continue collaborating without facing significant repercussions. For example, research misconduct involving influential scholars is often under the scanner on social media platforms (Garisto, 2023; Jack, 2023; Nesterak, 2023; Warren, 2023). This lack of awareness could prevent collaborators from making informed decisions about their continued collaboration.

Finally, our findings support prior research on the exoneration of researchers accused of scientific misconduct (Greitemeyer & Sagioglou, 2015). Stigmatized individuals might remain unaware of the extent of stigmatization surrounding their work or actions. In certain situations, collaborators might choose to give the benefit of the doubt to the authors who have faced retraction, as there is always a risk of inaccurate accusations (Greitemeyer & Sagioglou, 2015). This leniency could aid in the continuation of collaborative ties.

# 5   Conclusion

Overall, while we expected retractions to substantially impact the collaborative relationships of misconducting authors, our observations suggest a more nuanced and varied response within the scientific community. It's important to delve deeper into the specific dynamics of each case and consider broader contextual factors to understand why the anticipated consequences might not always materialize. Efforts to address fraud and enhance trust in scientific research should not overshadow the vast body of legitimate and valuable scientific research conducted by honest researchers. Building and maintaining trust in science is a collective responsibility shared by researchers, institutions, peer reviewers, and the wider scientific community.

Our findings have implications that go beyond the expected consequences. If scientific misconduct doesn't hinder researchers' collaborative progress, this raises a significant concern as it implies a lack of concern among researchers about ethics and integrity. In essence, mere collaboration does not guarantee research quality. Ensuring sound scientific norms within the community is crucial, given the strong link between scientific misconduct and individual understanding of integrity. Nations with high retraction rates must focus on enhancing research quality, fostering ethical behaviour, and providing research integrity training. To counteract repeated offences within scientific collaborations, creating a comprehensive legal framework for reporting misconduct and offering mentorship to institutions or research groups is recommended for overall research enhancement. On a worldwide scale, funding entities, publishing entities, and research communities should actively endorse responsible research conduct and promptly discredit any instances of "false science".

Institutional policies should encompass periodic mentoring sessions on research ethics, integrity, and their roles in advancing both science and society. Past research deals with the attrition rate of PhD scholars who were disillusioned with the misconduct among researchers (Mongeon & Larivière, 2016). During doctoral training, mandatory courses in research ethics should be integrated into the curriculum. For instance, India's University Grants Commission (UGC) in its 543rd meeting held on 9th August 2019 approved mandatory courses for awareness about publication ethics and publication misconduct (UGC, 2019).

Each research lab or mentor ought to engage in open discussions with their teams about the pitfalls of scientific misconduct, fostering ethical research and instilling a sense of social responsibility. Mentors and institutions should educate scholars on how to handle "failure" positively, providing skills that can benefit them throughout their lives. For instance, mentors could share their own experiences of unsuccessful experiments that didn't detrimentally affect their academic careers. To conclude, we suggest that institutions should offer professional counselling services, as supportive mentoring alone might not suffice in achieving this objective.

**Acknowledgements**

The authors received no funding for the research.